\documentclass[reqno,a4paper,12pt]{article}

\newcommand{\sect}[1]{\setcounter{equation}{0}\section{#1}\indent}

\usepackage{color}
\newcommand{\bref}[1]{(\ref{#1})}
\def\wG{{\widetilde{\mtG}}}
\usepackage{subeqnarray}
\def\half{{\textstyle\frac{1}{2}}}

\def\mtG{{\mathcal G}}

\def\mtP{{\mathcal P}}
\usepackage{stmaryrd}
\usepackage[centertags]{amsmath}
\usepackage[nosort]{cite}
\usepackage{array}
\usepackage{color}
\usepackage{amsfonts}
\usepackage{amssymb}
\usepackage{amsthm}
\usepackage{newlfont}
\begin{document}
\title{ Generalizations of Maxwell (super)algebras by the expansion method}
\vskip 2cm
\author{J. A. de Azc\'arraga$^a$, J. M. Izquierdo$^b$, \\
J. Lukierski$^c$ and M. Woronowicz$^c$}
\date{}
\maketitle
\begin{center}
$^a$  {\it Dept. Theor. Phys. and IFIC (CSIC-UVEG), Valencia Univ.
\newline
46100-Burjassot (Valencia), Spain}
\newline
$^b$  {\it Dept. Theor. Phys., Valladolid Univ., 47011-Valladolid, Spain}
 \newline
$^c$ {\it Institute of Theoretical Physics, Wroc{\l}aw Univ., 50-204 Wroc{\l}aw, Poland}
\newline
\date{\small October 3, 2012}
\end{center}
%
\begin{abstract}
The Lie algebras expansion method is used to show that the
four-dimensional spacetime Maxwell
(super)algebras and some of their generalizations can be derived in
a simple way as particular expansions of $o(3,2)$ and $osp(N|4)$.
\end{abstract}

\section{Introduction}
There are four methods of obtaining new Lie (super)algebras from a given
one: contractions, deformations, extensions and expansions. Contractions
and deformations lead to new algebras of the same dimension as the original one.
The same can be said of an extension $\wG$ of a Lie algebra $\mtG$ by another one
$\mathcal{A}$ (see {\it e.g.} \cite{hiazlu3}) in the sense that
 $\mathrm{dim} \wG = \mathrm{dim} \mtG + \mathrm{dim}\mathcal{A}$
since $\wG=\mtG/\mathcal{A}$. A fourth way of obtaining new Lie (super)algebras
from a given $\mathcal{G}$ ($s\mathcal{G}$) is the {\it expansion of algebras}, first used
in \cite{hiazlu4} and studied in general in \cite{hiazlu1,hiazlu2}; see \cite{IRP:06}
for further developments. In contrast with the
first three procedures, expanded algebras have, in general, higher dimension
than the original $\mathcal{G}$ because expansions give rise to additional generators
(expansions also include contractions as a particular case \cite{hiazlu1},
in which they are dimension-preserving). In this paper we shall consider
some basic aspects of the expansion procedure to
derive Maxwell (super)algebras and other new generalizations as expanded algebras.

The main idea of the Lie algebras expansion method is to promote the
standard Maurer-Cartan (MC) forms $\omega^i$ of the Lie algebra $\mtG$
of a group $G$ (resp. superalgebra and supergoup),
\begin{equation}
\label{eazlu1.1}
\theta (g)= g^{-1} \, dg = \omega^i X_i\ ,\
i=1,\dots, \textrm{dim}\, \mtG\ , \ X_i\in \mtG\ , \ g \in G \; ,
\end{equation}
to functions $\omega^i (\xi)$ of a real parameter $\xi$. We shall consider
here expansions of the $\omega$'s with the generic form \cite{hiazlu1}
\begin{equation}
\label{eazlu1.2}
\omega^{i_p} (\xi) = \sum\limits^{M_p}_{\alpha_p=p} \xi^{\alpha_p} \omega^{i_p,\alpha_p}=
\xi^p\omega^{i_p, p}+\xi^{p+1}\omega^{i_p,p+1}+\dots \omega^{i_p,M_p}\;,
\end{equation}
where $p$ refers to the subspace $V_p$ in the splitting
$\mtG=V_0\oplus V_1\oplus\dots V_p\oplus\dots$
of the Lie algebra vector space (thus, $i_p=1,\dots,\mathrm{dim}V_p$,
$\Sigma_p  \mathrm{dim} V_p=\mathrm{dim}\mtG$); $\alpha_p$ is the
power of $\xi$ in the series expansion that accompanies a
given $\omega^{i_p,\alpha_p}$, which is therefore characterized
by the values of the index $i_p$ and the order $\alpha_p$
in the expansion of $\omega^{i_p}$. Depending on the problem,
$\xi$ will be expressed as different powers
of $\lambda$ with dimensions [$\lambda$]=$L^{-\frac{1}{2}}$.
When $M_p=0$, $\omega^{i_p,0}=\omega^{i_p}$ {\it i.e.},
the $\omega$'s are the original MC forms for the various
subspaces $V_p\subset\mtG$. Certain terms of the expansion
may be absent in \bref{eazlu1.2}, as we will see later.

After inserting the expansions \bref{eazlu1.2} into the MC
equations for $\mtG$ we get
\begin{equation}
\label{eazlu1.3}
d\omega^i (\xi) = - \half\, c^i_{\, jk}\,\omega^j (\xi) \wedge
\omega^k (\xi) \, .
\end{equation}
Then, equating the coefficients of equal powers $\alpha_p$ of $\xi$
at the right and left hand sides, a set of equations for the
differentials $d\omega^{i_p, \alpha_p}$ is obtained. In principle,
the various $M_p$ in \eqref{eazlu1.2} could be arbitrarily large, but the idea of
the expansion method is to cut the series consistently {\it i.e.}, in such a way that the
retained $\{\omega^{i_p,\alpha_p}\}$ ($\alpha_p \leq M_p$) become the MC forms of a new,
{\it  expanded Lie (super)algebra} characterized by the MC equations satisfied by
these $\omega^{i_p,\alpha_p}$'s. The closure of $d$ on the set $\{\omega^{i_p,\alpha_p}\}$
requires that the highest $M_p$'s satisfy certain relations to guarantee that the
various expressions for $d\omega^{i_p,\alpha_p}$ define the MC equations of a
new Lie (super)algebra, the {\it expansion of $\mathcal{G}$}, denoted
$\mathcal{G}(M_0,\ldots,M_p)$. In this paper we shall only consider
splittings with $p=0,1,2$; only even or only odd powers of $\xi$ will appear in the expansions.

It is well known that the $D=4$\ Poincar\'e algebra
${\mtP}(3,1)=t_4{\niplus}o(3,1)$\ and the $D=4$\
Poincar\'e superalgebras $s{\mtP}(3,1|N)$\ $\ (N=0,1,2...)$ can be obtained, respectively, as
Wigner-\.In\"on\"u (WI) contraction of the $D=4$\ $adS$ algebra $o(3,2)$\ and of the
$D=4$\ $adS$ superalgebras $osp(N|4)$. As shown in \cite{hiazlu4,hiazlu1}
the expansion method goes beyond the original IW contraction of algebras
or its generalizations (as \cite{W-W:00}), all contractions being characterized
by  the equality of the dimensions of the original and the contracted
algebras. But, as it is the case with IW contractions (where the algebra
with respect to which the contraction is made is preserved in the process),
expanded algebras also keep memory of the original algebra
(through eqs.~\eqref{eazlu2.1},\eqref{eazlu2.1a} below). This fact is
particularly useful as a guiding principle to identify known algebras
as expansions or to generate new ones with some desired properties.
Moreover, since as mentioned contractions constitute a particular case
\cite{hiazlu1} of expansions, it is natural to use the expansion method
by taking the $o(3,2)$\ algebra and $osp(N|4)$\ superalgebras
as starting points in order to obtain various enlargements of the
$D=4$\ Poincar\'e algebra and the $D=4$\ Poincar\'e superalgebras.
We shall show that the expansion procedure leads, in particular, to a
16-dimensional enlargement of $D=4$\ Poincar\'e algebra containing six
additional tensorial Abelian charges, the Maxwell algebra\footnote{The name
`Maxwell' was proposed in \cite{hiazlu6}
for a sixteen generators algebra larger than the Bacry-Combe-Richard (BCR)
\cite{hiazlu5} one, which has  six generators plus two central charges.
The BCR algebra describes the symmetry of a relativistic particle
in a constant e.m. field.}
 (see e.g. \cite{hiazlu5,hiazlu6,Ne-Ol:00,hiazlu7,hiazlu8})
 and, further, to its generalizations.
Analogously, the expansion of supersymmetric $adS$\ algebra $osp(N|4)$\ will
produce known as well as new supersymmetric generalizations of Maxwell
algebra.

The plan of this paper is the following. Sect.~\ref{ExpGral} briefly
recalls some general aspects of the expansion procedure, with
particular attention to Lie algebras (and superalgebras)
with a symmetric coset structure. Sect.~\ref{LorMax} considers
the expansions of the $o(3,1)$ Lorentz algebra; further, using $o(3,2)$ and
the splitting $o(3,2)= o(3,1) \oplus \frac{o(3,2)}{o(3,1)}$, the Maxwell
algebra and its generalization are obtained as specific expansions
of $D=4$ $adS$ algebra $o(3,2)$. In Sect.~4 the expansion method is applied to a suitable
coset decomposition of the $D=4$ $adS$ superalgebra $osp(N|4)$; this will
lead to a new supersymmetric version of Maxwell algebra. We
conclude here by stressing that the expansion method is general, and
provides an effective algebraic scheme to derive larger (super)algebras from
a given one.

\sect{Expansions of Lie (super)algebras: general considerations}
\label{ExpGral}

In a rather general framework \cite{hiazlu1}, the MC equations for the
expansion $\mathcal{G}(M_0,\ldots,M_p)$ follow from eqs.~\eqref{eazlu1.2}
and \bref{eazlu1.3}. They have the form
\begin{eqnarray}
\label{eazlu2.1}
&&d\omega^{k_s,\alpha_s} = - \half \, C^{k_s,\alpha_s}_{\,i_p,\beta_p\  j_q,\gamma_q} \,
\omega^{i_p,\beta_p} \wedge
\omega^{j_q,\gamma_q}\; ,\\
\nonumber
i_{p,q,s}&=& 1,\dots, \mathrm{dim}\, V_{p,q,s}\ , \ \alpha_p, \beta_p,
\gamma_p =p, p+1, p+2, \dots, M_p \; ,
\end{eqnarray}
where the $M_p$ have to satisfy certain conditions and the structure constants
of the expansion $\mathcal{G}(M_0,\ldots,M_p)$ are given in terms of those
of $\mathcal{G}$ by
\begin{equation}
\label{eazlu2.1a}
    C^{k_s,\alpha_s}_{\, i_p,\beta_p\  j_q,\gamma_q} =
\left\{
        \begin{array}{ll}
           0, & \hbox{$\beta_p+\gamma_q\neq\alpha_s$} \\
           c^{k_s}{}_{\, i_p j_q}, & \hbox{$\beta_p+\gamma_q=\alpha_s$} \quad .
           \end{array}
        \right.
\end{equation}
We shall consider the following cases of the above general
structure:

1) All the MC forms $\omega^i$ in eq.~\bref{eazlu1.2} are expanded
similarly,
\begin{equation}
\label{eazlu2.2}
\omega^i (\xi) = \sum\limits^{M}_{\alpha=0} \xi^\alpha \, \omega^{i,\alpha} \;,\
i=1,\dots,\mathrm{dim}\mathcal{G}  \; ,
\end{equation}
{\it i.e.} $\mathcal{G}=V_0$, $i_0=i$. Eqs.~\bref{eazlu2.2}
in \bref{eazlu2.1} give
\begin{equation}
\label{eazlu2.3}
d\omega^{i,\alpha}= - \half \, c^i_{\, jk} \,
\sum\limits^{\alpha}_{\beta=0} \omega^{j,\beta} \wedge
\omega^{k,\alpha-\beta} \quad, \quad \alpha{=}0,1 \ldots M \; .
\end{equation}
The resulting Lie algebra expansions, denoted $\mathcal{G}(M)$, have generators
$\{X_{j,\beta}\}= ( X_{j, 0}, X_{j, 1} \ldots X_{j,M} )$ dual to
the MC forms $\{\omega^{i,\alpha}\}=(\omega^{i,0}, \omega^{i,1},\dots,\omega^{i,M})$
that satisfy the MC equations \bref{eazlu2.3}.
Consistency requires $d (d \omega^{i,\alpha}) \equiv 0$; this follows
using \bref{eazlu2.3} repeatedly for $d\omega^{j,\beta}$ etc.
in the $r.h.s.$ of $d (d \omega^{i,\alpha})=0$ and the Jacobi identity
(JI) for $\mtG$. Alternatively, $d (d \omega^{i,\alpha})\equiv 0$
follows as a consequence of the JI for $\mathcal{G}(M)$. The dimension
of the expansions $\mathcal{G}(M)$ is
dim$\,\mathcal{G}(M){=}(M+1)\times\mathrm{dim}\,\mathcal{G}$.
Eq.~\bref{eazlu2.3} implies
\begin{equation}
\label{eazlu2.4}
\left[ X_{j,\alpha} , X_{k,\beta}\right] =0 \qquad \hbox{if}\quad
\alpha+\beta > M \; .
\end{equation}
Therefore $\mathcal{G}\,(M)$ contains
$\ell$ sets of (dim$\,\mathcal{G}$)-dimensional abelian subalgebras of
generators  $\{X_{j, \ell+1}\} \ldots \{X_{j, M}\}$ when
$M=2\ell$ even, and $\ell$ sets $\{X_{j, \ell}\} \ldots  \{X_{j, M}\}$
in the odd $M{=}2\ell-1$ case.

To be consistent later with the notation in the supersymmetric
case it will prove convenient to set $\xi=\lambda^2$, $2M=N$
and relabel the expansion as $\mathcal{G}(N)$.
Then, eq.~\bref{eazlu2.2} reads
\begin{equation}
\label{eazlu2.2a}
  \omega^i (\lambda) = \sum\limits^{N}_{\alpha=0,\, \alpha\, \textrm{even}}
  \lambda^\alpha \, \omega^{i,\alpha} \;, i=1,\dots,\mathrm{dim}\mathcal{G} \; .
\end{equation}

2) Let us assume that the algebra has a symmetric coset structure,
$\mtG = \mathcal{H}\oplus\mathcal{K}$, with generators
$H_l \in \mathcal{H}$, $K_r \in \mathcal{K}$ so that the indices in
$\mathcal{H}$ ($\mathcal{K}$) take the values
$l,m,n{=}1 \ldots \mathrm{dim}\mathcal{H}$
($r,s{=}1\ldots \mathrm{dim}\mathcal{K}$). Then,
\begin{eqnarray}
\label{eazlu2.5}
&
\left[ H_l, H_m \right] = c^n{}_{lm} H_n
\quad , \quad
\left[ H_l, K_r \right] = c^s{}_{lr} K_s \quad ,
\nonumber\\ \\[-12pt]
&\left[ K_r, K_s \right] = c^l{}_{rs} H_l \; .
\nonumber
\end{eqnarray}
If we denote the dual MC forms of the subalgebra $\mathcal{H}$ by
$\omega^l$ ($\omega^l (H_m)=\delta^l_{\, m}$)
and $e^r$ are those of $\mathcal{K}$
($e^r (K_s)= \delta^r_s$) the algebra \bref{eazlu2.5}
is equally characterized by its MC equations,
\begin{eqnarray}
\label{eazlu2.6}
&
d\omega^l = - \half (\, c^l{}_{ mn} \, \omega^m \wedge
\omega^n  + c_{rs}^l e^r \wedge e^s )\; ,
\nonumber\\ \\[-12pt]
&
d e^r = - c^r{}_{ ms} \, \omega^m \wedge e^s \; .
\nonumber
\end{eqnarray}
Then, due to the symmetric coset structure, the expansions of the MC
forms take the form \cite{hiazlu1}
\begin{equation}
\label{eazlu2.7a}
\omega^l (\xi) =
\sum\limits^{M_0}_{\alpha_0=0,\, \alpha_0 \; \mathrm{even}} \xi^{\alpha_0} \, \omega^{l,\alpha_0} \quad ,
\quad e^r(\xi) =
\sum\limits^{M_1}_{\alpha_1=1,\, \alpha_1\; \mathrm{odd}} \xi^{\alpha_1} \, e^{r,\alpha_1} \quad ,
\end{equation}
{\it i.e.} $p=0$, $i_0=l$ and $p=1$, $i_1=r$ in eq.~\bref{eazlu1.2}
respectively. When  $M_1=M_0+1$ or $M_1=M_0-1$ \cite{hiazlu1} the retained forms
determine a Lie algebra. To compare these expressions with the supersymmetric ones
avoiding the odd powers in \eqref{eazlu2.7a}, we take as mentioned
$\xi=\lambda^2$, $2M_0=N_0$, $2M_1=N_2$. Then, eqs.~\eqref{eazlu2.7a} read
\begin{equation}
\label{eazlu2.7b}
\omega^l (\lambda) =
\sum\limits^{N_0}_{\alpha_0 =
0,\, \mathrm{mod}\, 4} \lambda^{\alpha_0} \, \omega^{l,\alpha_0} \quad ,\quad
e^r(\lambda) =
\sum\limits^{N_2}_{\alpha_2 =2,\, \mathrm{mod}\, 4}
\lambda^{\alpha_2} \, e^{r,\alpha_2}\ .
\end{equation}
We shall call $\mathcal{H}=V_0$ and $\mathcal{K}=V_2$ referring to the first
powers in $\lambda$ (rather than $\xi$) that appear in the expansion of the corresponding
MC forms, and denote the expansions $\mathcal{G}(N_0,N_2)$.
In Sect.~3 we will consider the case $M_0=2$, $M_1=1$, {\it i.e.}
$N_0{=}4$ and $N_2{=}2$; this will lead to the Maxwell algebra as the expansion
$o(3,2)(N_0=4,N_2=2)$

3) The case of superalgebras corresponds to adding the
fermionic sector $V_1$ of generators $F_\alpha \in V_1$
to the bosonic one $\mtG = V_0\oplus V_2$, for which we still assume
eqs.~\eqref{eazlu2.5}. Thus, $s \mtG = V_0 \oplus V_1 \oplus V_2$
and the commutation relations of $s\mtG$ are given by
eqs.~\eqref{eazlu2.5} plus
\begin{eqnarray}
\label{eazlu2.8}
&&
\{ F_\alpha , F_\beta \} = c^n{}_{\alpha\beta} H_n + c^r{}_{\alpha\beta}  K_r \; ,
 \nonumber\\ \\[-12pt]
\cr
&&
[H_l , F_\alpha ] = c^\beta{}_{l\alpha}  F_\beta \quad , \quad
[K_r , F_\alpha ] = c^\beta{}_{r\alpha}  F_\beta  \;,
\nonumber
\end{eqnarray}
where $\alpha,\beta {=}1 \ldots \mathrm{dim}\, V_1$ above
refer to the spinorial index of the fermionic generator.
Introducing fermionic MC forms
$\psi^\alpha$, $\psi^\alpha (F_\beta)=\delta^{\alpha}_{\,\beta}$,
the MC equations for the Lie superalgebra $s\mtG$ follow from
\eqref{eazlu2.5}, \bref{eazlu2.8},
\begin{eqnarray}
\label{eazlu2.9}
&&
d\omega^l = - \half (c^l_{\, mn} \omega^m \wedge \omega^n
+ c_{rs}^l e^r \wedge e^s + c^l_{\alpha\beta} \psi^\alpha \wedge \psi^\beta) \; ,
\nonumber\\[8pt]
&&
de^r = - \half (2c^r_{ms} \omega^m \wedge e^s + c^r_{\, \alpha\beta}
\psi^\alpha \wedge \psi^\beta ) \; ,
\\[8pt]
&&
d\psi^\alpha = -  (
c^\alpha_{\, \beta l} \psi^\beta \wedge \omega^l
+ c^\alpha_{\, \beta r} \psi^\beta \wedge e^r ) \; .
\nonumber
\end{eqnarray}

Given the coset structure of the superalgebra $s\mtG$  ($p=0,1,2$),
we take $\xi=\lambda$ so that $\lambda$ accompanies
the first term in the expansion
\begin{equation}
\label{eazlu2.10}
\psi^\alpha (\lambda) =
\sum\limits^{N_1}_{\alpha_1=1, \, \alpha_1\; \mathrm{odd}}
\lambda^{\alpha_1} \psi^{\alpha, \alpha_1} \; .
\end{equation}
Note that the first index $\alpha$ in $\psi^{\alpha, \alpha_1}$ is the
usual spinorial one and that the second index $\alpha_1$ refers
to the order (power) of the expansion of the fermionic MC form
$\psi^\alpha$  of $s\mtG$. The expansions of the bosonic MC forms
are accordingly
\begin{equation}
\label{eazlu2.7}
\omega^l (\lambda) =
\sum\limits^{N_0}_{\alpha_0=0,\, \alpha_0\, \mathrm{even}} \lambda^{\alpha_0} \, \omega^{l,\alpha_0}
\quad ,\quad e^r(\lambda) =
\sum\limits^{N_2}_{\alpha_2=2,\, \alpha_2\, \mathrm{even}} \lambda^{\alpha_2} \, e^{r,\alpha_2}\ .
\end{equation}
The powers of $\lambda$ in eqs.~\eqref{eazlu2.10}, \eqref{eazlu2.7},
where [$\lambda]=L^{-\frac{1}{2}}$, will determine later suitable physical dimensions
for the MC forms in the expansion and for their dual Lie (super)algebra generators.
In fact, if the MC forms are identified with the one-form
fields of a physical theory, the lower orders will lead to the standard physical
dimensions of the bosonic and fermionic fields in geometrized units ({\it e.g.},
the one-form $\psi^{\alpha,1}$  in eq.~\eqref{eazlu2.10} has
dimension $[\psi^{\alpha,1}]=L^{\frac{1}{2}}$, which gives
$[Q_\alpha]=L^{-\frac{1}{2}}$ for its dual generator in the
superalgebra, etc). It is also seen that supersymmetry makes the expansion of the bosonic
part to be of the form \bref{eazlu2.7} rather than \bref{eazlu2.7b}.
If we now set the fermionic sector (eq.~\eqref{eazlu2.10}) in the
expansion of the superalgebra  $s\mtG$ equal to zero, the result gives a
possible expansion of the bosonic subalgebra $V_0\oplus V_2 \subset s\mtG$,
which is consistent and different from the one in eq.~\bref{eazlu2.7b}.

In Sec.~\ref{LorMax} we shall consider the bosonic expansion given by eq.~\eqref{eazlu2.7}
for $N_0=4=N_2$ to derive a new generalization of Maxwell algebra, and in Sect.~\ref{sMax} we
shall consider the supersymmetric expansion (eqs.~\eqref{eazlu2.10}, \eqref{eazlu2.7})
with $N_0=4{=}N_2$ and $N_1=3$ (case (a) below) to obtain new supersymmetrizations
of the Maxwell algebra. Note that a set of integers $(N_0,N_1,N_2)$ will
not lead to an expansion $s\mathcal{G}(N_0,N_1,N_2)$
unless one of the following conditions is satisfied: (a) $N_0=N_1+1=N_2$,
(b) $N_0=N_1-1=N_2$, (c) $N_0=N_1-1=N_2-2$ or (d) $N_0-2=N_1-1=N_2$.
Case (d) would be absent if we allowed for an $H$ component in the
second commutator of \bref{eazlu2.5} (see \cite{hiazlu1,hiazlu4}
for details).

\sect{Generalized $\bold{D=4}$ Lorentz and Maxwell algebras as expansions}
\label{LorMax}

\noindent
1) \textsl{Expansions of the $D{=}4$ Lorentz algebra}.

The $o(3,1)$ Lorentz algebra is given by the relations
 ($\mu,\nu=0,1,2,3$).
 \begin{equation}
 \label{eazlu3.1}
 [M_{\mu\nu}, M_{\rho\sigma}] =
 (\eta_{\rho\nu}M_{\mu\sigma} -
 \eta_{\sigma\nu} M_{\mu\rho} ) -(\mu \leftrightarrow \nu)
 \end{equation}
 or, alternatively, by the MC equations for the Lorentz algebra which,
 with $\omega^{\mu\nu} (M_{\rho\tau})=\delta^{\mu\nu}_{\rho\tau}$, are
 \begin{equation}
 \label{eazlu3.2}
 d\omega^{\mu\nu} = - \, \omega^\mu{}_{\, \rho} \wedge \omega^{\rho \nu} \quad .
 \end{equation}
 We now use the expansion formula \bref{eazlu2.2a} with $N{=}4$,
 \begin{equation}
 \label{eazlu3.3}
 \omega^{\mu\nu} (\lambda) = \omega^{\mu\nu ,0} + \lambda^2\, \omega^{\mu\nu ,2}
 + \lambda^4 \omega^{\mu\nu, 4} \; .
 \end{equation}
 From \bref{eazlu2.3} and taking into account that
 only even orders of $\alpha$ and $\beta$ appear,  we obtain the relations
 \begin{subeqnarray}
 &&
 d\,\omega^{\mu\nu, 0} = -  \omega^\mu{}_\rho{}^{, 0} \wedge \omega^{\rho\nu, 0} \; ,
 \\[8pt]
  \label{eazlu3.4a}
 &&
 d\,\omega^{\mu\nu, 2} = -
 (
  \omega^\mu{}_\rho{}^{, 0} \wedge \omega^{\rho\nu, 2} +
  \omega^\mu{}_\rho{}^{, 2} \wedge \omega^{\rho\nu, 0}  ) \; ,
  \label{eazlu3.4b}
 \\[8pt]
 &&
  d\,\omega^{\mu\nu, 4} = -
  (
  \omega^\mu{}_\rho{}^{, 0} \wedge \omega^{\rho\nu, 4} +
  \omega^\mu{}_\rho{}^{, 2} \wedge \omega^{\rho\nu, 2} +
  \omega^\mu{}_\rho{}^{, 4} \wedge \omega^{\rho\nu, 0} )  \; .
  \label{eazlu3.4c}
 \end{subeqnarray}

 Eqs. (\ref{eazlu3.4a}a--c) constitute the MC equations for the
 ($6{\times}(2{+}1$))-dimensional Lie-algebra expansion $o(3,1)(N=4)$.
 Introducing the generators $M_{\mu\nu}$, ${\widetilde{Z}}_{\mu\nu}$,
 $Z_{\mu\nu}$ dual respectively to $\omega^{\mu\nu, 0}$, $\omega^{\mu\nu, 2}$,
 $\omega^{\mu\nu, 4}$ we obtain, besides \bref{eazlu3.1}, the following
 $o(3,1)(4)$ commutators (we usually omit the vanishing ones)
 \begin{subeqnarray}
 \label{eazlu3.5a}
 &&
 [M_{\mu\nu}, {\widetilde{Z}}_{\rho\sigma} ] =
 (\eta_{\rho\nu}\, {\widetilde{Z}}_{\mu \sigma}
 - \eta_{\sigma \nu} {\widetilde{Z}}_{\mu\rho})- (\mu\leftrightarrow \nu) \; ,
 \\[8pt]
 &&
 [M_{\mu\nu}, {{Z}}_{\rho\sigma} ] =
 (\eta_{\rho\nu}\, {{Z}}_{\mu \sigma}
 - \eta_{\sigma \nu} {{Z}}_{\mu\rho}) - (\mu\leftrightarrow \nu)  \; ,
  \label{eazlu3.5b}
 \\[8pt]
 &&
 [ {\widetilde{Z}}_{\mu\nu}, {\widetilde{Z}}_{\rho\sigma} ] =
 (\eta_{\rho\nu}\, {{Z}}_{\mu \sigma}
 - \eta_{\sigma \nu} {{Z}}_{\mu\rho}) - (\mu\leftrightarrow \nu) \; ;
 \label{eazlu3.5c}
 \end{subeqnarray}
 in agreement with the general relations \bref{eazlu2.4} we also get
 $[ Z_{\mu\nu}, Z_{\rho\tau} ]=0$.
 Therefore, $o(3,1)(4)$ is the semidirect product
 of the Lorentz algebra and the  ideal generated by
 (${\widetilde{Z}}_{\mu\nu}, Z_{\mu\nu}$) in which
 the generators $Z_{\mu\nu}$ are central.
\vskip .2cm

\noindent
 2) \textsl{$D=4$ Maxwell algebra}.

  Let us now consider $D=4$ $adS$ algebra $o(3,2)$ with the
 splitting \bref{eazlu2.5}, namely $o(3,2){=}o(3,1)\oplus \frac{o(3,2)}{o(3,1)} \equiv V_0 \oplus V_2$.
 Denoting the generators in the coset $\frac{o(3,2)}{o(3,1)}$ describing
 the curved translations in $D{=}4$ $AdS$ space by ${\mtP}_\mu$, the algebra $o(3,2)$
 is obtained supplementing the Lorentz algebra \bref{eazlu3.1} with
 \begin{equation}
  \label{eazlu3.6}
  \left[ M_{\mu\nu}, {\mtP}_\rho \right]  =
  \,2( {\mtP}_{\mu }\eta_{\nu\rho} - {\mtP}_{\nu }\eta_{\mu\rho}) \quad, \quad
  [ {\mtP}_\mu, {\mtP}_\nu ]   = \, M_{\mu\nu} \quad .
\end{equation}
 The $o(3,2)$ algebra MC equations are
\begin{equation}
\label{eazlu3.8a}
d\omega^{\mu\nu} =  -  \,\omega^\mu{}_\rho
 \wedge \omega^{\rho\nu} -  e^\mu \wedge e^\nu  \quad, \quad
d\, e^\mu =  -  \omega^\mu_{}\rho \wedge e^\rho \; ,
\end{equation}
where all the generators in \eqref{eazlu3.6} and the forms in
\eqref{eazlu3.8a} are dimensionless. The well
known contraction of $o(3,2)$ algebra to the Poincar\'e algebra
is obtained as an expansion if we use expression
\bref{eazlu2.7a} and retain there only the first terms $(M_0=0,M_1=1)$.

   Let us further consider the expansion using the form
\bref{eazlu2.7b} for the coset structure \bref{eazlu2.5}
(with $V_0=o(3,1)$ and $V_2= \frac{o(3,2)}{o(3,1)}$)
with $N_0=4$, $N_2=2$,
 \begin{equation}
 \label{eazlu3.12}
 \omega^{\mu\nu} (\lambda) = \omega^{\mu\nu, 0} + \lambda^4\, \omega^{\mu\nu , 4}
 \quad, \quad
e^\mu (\lambda) = \lambda^2 \, e^{\mu , 2} \; .
 \end{equation}
Inserting these expressions in the $o(3,2)$  MC equations above,
and identifying the coefficients of equal powers of $\lambda$, we obtain
\begin{eqnarray}
\label{eazlu3.99}
&&
\nonumber
d\,\omega^{\mu\nu, 0} = -  \omega^\mu{}_\rho{}^{, 0} \wedge \omega^{\rho\nu, 0}
\; , \quad
 d\, e^{\mu, 2} = -  \omega^\mu{}_\nu{}^{, 0} \wedge
 e^{\nu, 2}  \ ,
\\ [3pt]
&& d\,\omega^{\mu\nu, 4} = -
  (
  \omega^\mu{}_\rho{}^{, 0} \wedge \omega^{\rho\nu, 4} +
  \omega^\mu{}_\rho{}^{, 4} \wedge \omega^{\rho\nu, 0}
+  e^{\mu, 2} \wedge e^{\nu, 2})  \; .
 \end{eqnarray}
These are the MC equations of the {\it Maxwell algebra} as the
expansion $o(3,2)(N_0=4,N_2=2)$. In terms of the generators
$M_{\mu \nu}, P_\mu$ and $Z_{\mu \nu}$ dual to $\omega^{\mu \nu, 0}, e^{\mu ,2}$
and $\omega^{\mu \nu, 4}$ respectively, the commutators of the algebra are
\begin{eqnarray}
\label{eazlu3.111}
&[M_{\mu\nu}, M_{\rho\sigma}]& =
(\eta_{\rho\nu}M_{\mu\sigma} -
\eta_{\sigma\nu} M_{\mu\rho} ) -(\mu \leftrightarrow \nu) \; ,
\nonumber \\[3pt]
&[ P_\mu, P_\nu ] &=  Z_{\mu\nu}  \; ,
\nonumber \\[3pt]
&[ M_{\mu\nu}, P_\rho ] &=  \, 2(P_{\mu}\eta_{\nu \rho} - P_{\nu}\eta_{\mu \rho} )  \;  ,
\end{eqnarray}
plus eq.~(\ref{eazlu3.5b}b).
\vskip .2cm

\noindent
3) \textsl{The generalized $D=4$ Maxwell algebra}.

 To obtain it, now expand the MC forms $(\omega^{\mu\nu}, e^\mu)$ dual to
$(M_{\mu\nu}, {\mtP}_{\mu})$  as in eq.~\bref{eazlu2.7} with
$N_0{=}4$ and $N_2{=}4$ (these expansions follow, as shown in case 3
in Sec.~\ref{ExpGral}, from those of $s\mathcal{G}$).
Besides eq.~\bref{eazlu3.3} we have from eq.~\eqref{eazlu2.7}
\begin{equation}
 \label{eazlu3.7}
 e^\mu (\lambda) = \lambda^2\, e^{\mu, 2} + \lambda^4 \, e^{\mu, 4}  \; .
 \end{equation}
Using the expansion \bref{eazlu3.3} for $\omega^{\mu\nu}$
and \bref{eazlu3.7} for $e^\mu$ in the $o(3,2)$ MC equations
\eqref{eazlu3.8a}, we obtain  eqs.~(\ref{eazlu3.4a}a--b),
the following modified eq.~(\ref{eazlu3.4c}c)
 \begin{equation}
 \label{eazlu3.9}
 d\,\omega^{\mu\nu, 4} = -
  (
  \omega^\mu{}_\rho{}^{, 0} \wedge \omega^{\rho\nu, 4} +
  \omega^\mu{}_\rho{}^{, 2} \wedge \omega^{\rho\nu, 2} +
  \omega^\mu{}_\rho{}^{, 4} \wedge \omega^{\rho\nu, 0} +
  e^{\mu, 2} \wedge e^{\nu, 2})
\end{equation}
 and
 \begin{equation}
 \label{eazlu3.10}
 d\, e^{\mu, 2} = -  \omega^\mu{}_\nu{}^{, 0} \wedge
 e^{\nu, 2} \quad , \quad
  d\, e^{\mu, 4} = - \, \omega^\mu{}_\nu{}^{, 0} \wedge
 e^{\nu, 4} - \, \omega^\mu{}_\nu{}^{, 2} \wedge
 e^{\nu, 2} \; .
 \end{equation}
 Introducing the generators $P_\mu, Z_\mu$ dual to $e^{\mu , 2}, e^{\mu , 4}$,
 it follows from eqs.~(\ref{eazlu3.9}), (\ref{eazlu3.10})
 that the expanded algebra, denoted
 $o(3,2)(4,4)$, provides the generalization of the Maxwell
 algebra given by the eqs.~\eqref{eazlu3.111}, (\ref{eazlu3.5a}a--c), plus
\begin{subeqnarray}
 \label{eazlu3.11a}
[ M_{\mu\nu}, Z_\rho ] &=&  \,  2(Z_{\mu}\eta_{\nu \rho} - Z_{\nu}\eta_{\mu \rho} )  \; ,
 \\[3pt]
\label{eazlu3.11d}
[{\widetilde{Z}}_{\mu\nu}, P_\rho ] &=&  \, 2(Z_{\mu}\eta_{\nu \rho} - Z_{\nu}\eta_{\mu \rho} ) \; ,
 \\[3pt]
  \label{eazlu3.11e}
0=[{{Z}}_{\mu\nu}, Z_{\rho\tau} ] =
[ {{Z}}_{\mu\nu}, {\widetilde{Z}}_{\rho\tau} ] =[ {{Z}}_{\mu}, Z_\nu ]&=&
[Z_\rho, \widetilde{Z}_{\mu \nu}]=[Z_\rho, Z_{\mu \nu}] \; .
  \end{subeqnarray}

Thus, the expansion $o(3,2)(4,4)$ contains the Maxwell algebra
 as the subalgebra generated by ($M_{\mu\nu}, P_\nu, Z_{\mu\nu}$).
 The addition of ${\widetilde{Z}}_{\mu\nu}$ provides, through
 eq.~(\ref{eazlu3.5a}c), the `bosonic roots' of the abelian charges
  $Z_{\mu\nu}$ appearing in the Maxwell algebra.
 The abelian vector charges $Z_\rho$, dual to $e^{\mu,4}$ in
 eq.~\eqref{eazlu3.7}, are central but for their Lorentz vector
 character.

 \sect{$\boldsymbol N$-extended Maxwell $\bold{ D=4}$ superalgebras as expansions of
 $\boldsymbol{osp(N|4)}$}
\label{sMax}

To obtain the $N$-extended $D=4$ Maxwell superalgebras we now
expand the $D=4$ $adS$ superalgebra $osp(N|4)$ with the coset splitting
of case 3) in Sec.~\ref{ExpGral}. Explicitly,
$osp(N|4)=V_0\oplus V_2\oplus V_1$ is given by $(N=1,2,3,\dots)$
  \begin{equation}
  \label{eazlu4.1}
osp(N|4) = [o(1,3)\oplus o(N) ] \oplus
  \frac{sp(4)}{o(1,3)} \oplus
  \frac{osp(N|4)}{sp(4)\oplus o(N)} \quad .
  \end{equation}
Since $sp(4)\simeq o(3,2)$ the algebra \bref{eazlu4.1} is the supersymmetric
counterpart of the $D{=}4$ $AdS$ algebra obtained by adding the
$\frac{N(N-1)}{2}$  generators $T^{ab}$ of $o(N)$ and $N$ real
$D{=}4$ Majorana spinor fermionic generators $\mathcal{Q}^a_\alpha $ ($a=1\ldots N$).
They satisfy the relations ($C=\gamma_0$ in the Majorana realization)
  \begin{eqnarray}
  \label{eazlu4.2}
  \{ \mathcal{Q}^a_\alpha, \mathcal{Q}^b_{\beta} \} =
  \delta^{ab}  (C\gamma^\mu)_{\alpha\beta} {\mtP}_\mu
  -\frac{1}{2} \delta^{ab} (C\gamma^{\mu\nu})_{\alpha\beta} \, M_{\mu\nu}
  + C_ {\alpha\beta}\, T^{ab} \; ,
  \end{eqnarray}
\begin{equation}
\label{eazlu4.3}
[ M_{\mu\nu}, \mathcal{Q}^a_\alpha ] =  (\mathcal{Q}^a \, \gamma_{\mu\nu})_\alpha\ ,\
[ {\mtP}_{\mu}, \mathcal{Q}^a_\alpha ] =  \half (\mathcal{Q}^a \, \gamma_{\mu})_\alpha\ ,\
[ T^{ab}, \mathcal{Q}^c_\alpha ] =
2( \mathcal{Q}^{a}_{\alpha}  \delta^{bc} - \mathcal{Q}^{b}_{\alpha}  \delta^{ac} ) \; ,
\end{equation}
where $T_{ab}$ are the internal symmetry $o(N)$ generators
\begin{equation}
\label{eazlu4.4}
[ T_{ab }, T_{cd} ] =  (
\delta_{cb} T_{ad} - \delta_{db} T_{ac}) - (a\leftrightarrow b)  ) \quad , \quad a,b=1\ldots,N \; .
\end{equation}

Let $\omega^{ab}$ be the one-forms dual to $T_{ab}$ (the indices $a,b$,
being euclidean, can be placed up or down) and $\psi^\alpha_{a}$ the
fermionic MC forms dual to $Q^a_\alpha$. The splitting \bref{eazlu4.1} corresponds
to the following assignments of the MC one-forms in the generic
MC equations \bref{eazlu2.9}
\begin{equation}
\label{eazlu4.5}
\omega^l \rightarrow (\omega^{\mu\nu} , \omega^{ab}) \quad, \quad
e^r \rightarrow (e^\mu ) \quad, \quad \psi^\alpha \rightarrow (\psi^\alpha_{\, a}) \quad .
\end{equation}
For $d\omega^{\mu\nu}$, $de^\mu$ we obtain
($\overline{\psi}= \psi^{T}C$)
\begin{eqnarray}
\label{eazlu4.6}
d\,\omega^{\mu\nu} & = & -  \, \omega^{\mu}{}_\rho \wedge
\omega^{\rho\nu} - e^{\mu} \wedge e^{\nu}
 +\frac{1}{2} {\overline{\psi}}{}^{\alpha}_{\, a} (\gamma^{\mu\nu})_{\alpha\beta}
\psi^{\beta}_{\, a} \quad ,
\nonumber \\ \\[-12pt]
d\, e^{\mu} & = & -   \omega^{\mu}{}_\rho \wedge
e^{\rho}
- \frac{1}{2}  \,
{\overline{\psi}}{}^{\alpha}_{\, a} (\gamma^{\mu})_{\alpha\beta}
\wedge\psi^{\beta}_{\, {a}} \quad .
\nonumber
\end{eqnarray}
The $osp(N|4)$ MC equations require adding to \eqref{eazlu4.6}
those for $d\omega_{ab}$ and $d\psi^{\alpha}{}_a$,
\begin{eqnarray}
\label{eazlu4.7}
d\, \omega_{ab} & = & - \omega_{ac} \wedge \omega^c{}_b
- {\overline{\psi}}^{\alpha}{}_a \wedge\psi_{\alpha b} \quad ,
 \\  \nonumber \\[-12pt]
d\, \psi^{\alpha}{}_a & = & - \frac{1}{4} ( \omega^{\mu\nu} \gamma_{\mu\nu} )^\alpha{}_\beta
\wedge\psi^\beta{}_a
- \frac{1}{2} e^\mu \gamma_\mu{}^\alpha{}_\beta
\wedge\psi^\beta{}_a
+  \omega^b{}_a \wedge \psi^\alpha{}_b  \quad .
\label{eazlu4.8}
\end{eqnarray}
At this stage, all the above generators and MC forms are dimensionless.

To expand now $osp(N|4)$ we use the splitting \eqref{eazlu4.1} and
eqs.~\eqref{eazlu2.7}, \eqref{eazlu2.10} for $\omega^{\mu\nu}$, $\omega^{ab}$,
$e^\nu$ and $\psi^\alpha{}_i$, and choose $N_0{=}4{=}N_2$ in \bref{eazlu2.7}
and $N_1{=}3$ in \bref{eazlu2.10} (case a) at the end of Sec.~\ref{ExpGral}).
In all, we have
\begin{eqnarray}
\label{eazlu4.9}
\omega^{\mu\nu} (\lambda) & = & \omega^{\mu\nu, 0}
+ \lambda^2\, \omega^{\mu\nu ,2} +
\lambda^4 \, \omega^{\mu\nu ,4} \; ,
\nonumber
\\[3pt]
e^\mu (\lambda) & = &
\lambda^2\, e^{\mu , 2}
+ \lambda^{4} \, e^{\mu , 4} \; ,
\end{eqnarray}
\begin{eqnarray}
\label{eazlu4.10}
\omega^{ab} (\lambda) & = & \omega^{ab, 0}
+ \lambda^2\, \omega^{ab, 2} +
\lambda^4 \, \omega^{ab, 4} \; ,
\\[3pt]
\psi^{\alpha}{}_{a} (\lambda) & = &
\lambda \, \psi^{\alpha}{}_a{}^{,1}
+ \lambda^3 \, \psi^{\alpha}{}_a{}^{,3}  \; .
\label{eazlu4.11}
\end{eqnarray}
Now, inserting expressions \bref{eazlu4.9}-\bref{eazlu4.11} into
the $osp(N|4)$ MC equations \bref{eazlu4.6}-\bref{eazlu4.8}, we
obtain the MC equations of the expansion $osp(N|4)(N_0,N_1,N_2)=
osp(4|N)(4,3,4)$ plus other new superalgebras through the
following consecutive steps:
\vskip .2cm

i) The MC equations for $\omega^{\mu\nu, 0}$, $e^{\mu , 2}$, $\omega^{ab , 0}$,
$\psi^{\alpha, 1}$ already determine a superalgebra,
the expansion $osp(N|4)(0,1,2)$. This is \cite{hiazlu1},
in fact, a contracted algebra (in the generalized sense of Weimar-Woods \cite{W-W:00})
and hence has the same dimension as $osp(N|4)$.
Its MC equations are given by
\begin{eqnarray}
\label{eazlu4.12}
d\,\omega^{\mu\nu, 0} & = & -  \omega^{\mu}{}_{\rho}{}^{,0} \wedge
\omega^{\rho\nu , 0}   \; ,
\nonumber \\
\nonumber
d\, e^{\mu, 2} & = & -  \omega^{\mu}{}_{\rho}{}^{,0} \wedge
e^{\rho, 2}
-\frac{1}{2}  \,
{\overline{\psi}}{}^{\alpha}_{\, a}{}^{,1} (\gamma^{\mu})_{\alpha\beta}
\psi^{\beta}_{\, {a}}{}^{,1} \; ,
\\
d\, \omega_{ab , 0} & = & -  \omega_{ac , 0} \wedge \omega^c{}_b{}^{,0} \; ,
\nonumber \\
d\, \psi^{\alpha,1}{}_a & = & - \frac{1}{4} ( \omega^{\mu\nu , 0} \gamma_{\mu\nu} )^\alpha{}_\beta \,
\psi^\beta{}_a^{,1}
+  \omega^b{}_a^{,0} \, \psi^\alpha{}_b^{,1} \; .
\end{eqnarray}
After using the duality relations for $\omega^{\mu\nu}$, $e^\mu$ from Sect.~\ref{LorMax}
and $\omega^{ab,0} (T_{cd})=\delta^{ab}_{cd}$,
$\psi^{\alpha,1}{}_{a} (Q^b{}_\beta)=\delta^\alpha_{\beta} \delta^{a}_{b}$
one gets from \bref{eazlu4.12} that $osp(N|4)(0,1,2)=s{\mtP}^{(N)}$ {\it i.e.},
the standard $N$-extended Poincar\'{e} superalgebra
generated by ($M_{\mu\nu},Q_\alpha{}^{a}, P_\mu,  T^{ab}$) with
$[Q_\alpha{}^{a}]=L^{-\frac{1}{2}}$ and $[P_\mu]=L^{-1}$ as usual.
\vskip .2cm

ii) The remaining equations in  the expansion of $osp(N|4)$ provide an enlargement of
$s{\mtP}^{(N)}$ superalgebra, with new generators
${\widetilde{Z}}_{\mu\nu}$,
${{Z}}_{\mu\nu}$, $Z_{\mu}$, ${\widetilde{Y}}_{ab}$,
${{Y}}_{ab}$ and $\Sigma_{\alpha}{}^{b}$ dual, respectively, to the one-forms
$\omega^{\mu\nu ,2}$, $\omega^{\mu\nu ,4}$, $e^{\mu ,4}$, $\omega^{ab ,2}$, $\omega^{ab ,4}$
and $\psi^{\alpha,3}{}_a$. These MC forms have dimensions
$[\omega^{\mu\nu ,2}]=L$, $[\omega^{\mu\nu ,4}]=L^2$, $[e^{\mu ,4}]=L^2$,
$[\omega^{ab ,2}]=L$, $[\omega^{ab ,4}]=L^2$ and $[\psi^{\alpha,3}{}_a]=L^{\frac{3}{2}}$,
inverse, respectively, to those of the corresponding expanded Lie algebra
generators.

From \bref{eazlu4.6}-\bref{eazlu4.8} the $osp(N|4) (4,3,4)$ MC equations follow:
\begin{eqnarray}
\label{eazlu4.13}
d\,\omega^{\mu\nu, 2} & = & -  \omega^{\mu}{}_{\rho}{}^{,0} \wedge
\omega^{\rho\nu , 2} -  \omega^{\mu}{}_{\rho}{}^{,2} \wedge
\omega^{\rho\nu , 0}  +
 \frac{1}{2}{\overline{\psi}}{}^{\alpha,1}_{\, a} (\gamma^{\mu\nu})_{\alpha\beta}
\wedge\psi^{\beta}_{\, a}{}^{,1}\\
d\,\omega^{\mu\nu, 4} & = & - \omega^{\mu}{}_{\rho}{}^{,0} \wedge
\omega^{\rho\nu , 4} -  \omega^{\mu}{}_{\rho}{}^{,4} \wedge
\omega^{\rho\nu , 0} -  \omega^{\mu}{}_{\rho}{}^{,2} \wedge
\omega^{\rho\nu , 2}-  e^{\mu , 2} \wedge e^{\nu ,2} \nonumber
\\
&& +\frac{1}{2} {\overline{\psi}}{}^{\alpha,1}_{\, a} (\gamma^{\mu\nu})_{\alpha\beta}
\wedge\psi^{\beta,3}_{\, a} +
 \frac{1}{2} {\overline{\psi}}{}^{\alpha}_{\, a}{}^{,3} (\gamma^{\mu\nu})_{\alpha\beta}
\wedge\psi^{\beta,1}_{\, a}
\label{eazlu4.14}
\\
d\, e^{\mu ,4} & = & -  \omega^{\mu}{}_{\rho}{}^{,0} \wedge
e^{\rho ,4} -  \omega^{\mu}{}_{\rho}{}^{,2} \wedge
e^{\rho ,2} \nonumber
\\
&& - \frac{1}{2}  \,
{\overline{\psi}}{}^{\alpha,1}_{\, a} (\gamma^{\mu})_{\alpha\beta}
\wedge\psi^{\beta,3}_{\, {a}} - \frac{1}{2}   \,
{\overline{\psi}}{}^{\alpha,3}_{\, a} (\gamma^{\mu})_{\alpha\beta}
\wedge\psi^{\beta,1}_{\, {a}}
\label{eazlu4.15}
\\
d\, \omega_{ab}{}^{, 2} & = & -  \omega_{ac}^{,0} \wedge \omega^c{}_b{}^{,2}
-  \omega_{ac}^{,2} \wedge \omega^c{}_b{}^{,0}
- {\overline{\psi}}^{\alpha,1}{}_a \wedge \psi_{\alpha b}{}^{,1}
\label{eazlu4.16}
\\
d\, \omega_{ab}{}^{, 4} & = & -  \omega_{ac}^{,0} \wedge \omega^c{}_b{}^{,4}
- \omega_{ac}^{,4} \wedge \omega^c{}_b{}^{,0} -  \omega_{ac,2} \wedge \omega^c{}_b{}^{,2}
\nonumber
\\
&&
- {\overline{\psi}}^{\alpha,1}{}_a \wedge \psi_{\alpha b}{}^{,3}
- {\overline{\psi}}^{\alpha,3}{}_a \wedge \psi_{\alpha b}{}^{,1}
\label{eazlu4.17}
\\
d\, \psi^{\alpha,3}{}_a & = &
- \frac{1}{4} ( \omega^{\mu\nu ,0} \gamma_{\mu\nu} )^\alpha{}_\beta \wedge\psi^{\beta,3}{}_a
- \frac{1}{4} ( \omega^{\mu\nu ,2} \gamma_{\mu\nu} )^\alpha{}_\beta \wedge\psi^{\beta,1}{}_a
- \frac{1}{2} ( e^{\mu, 2} \gamma_\mu )^\alpha{}_\beta \wedge\psi^{\beta,1}{}_a\nonumber\\
&&
+ \omega^b{}_a{}^{,0} \wedge \psi^{\alpha,3}_b +  \omega^b{}_a{}^{,2} \wedge \psi^{\alpha,1}{}_b
\phantom{xxxxx}
\label{eazlu4.18}
\end{eqnarray}
Some terms in the above expressions may be added up, but we have
preferred to exhibit the way they are generated by the expansion.

\vskip .2cm
The expansion procedure leads to the possible
generalizations of $s{\mtP} (N)$ described below:
\vskip .2cm

\noindent
1) {\sl An extension of the $\{Q^a_\alpha,Q^b_\beta \}$ anticommutator of the $N$-extended
superPoincar\'e algebra by the abelian $SO(1,3)$ tensorial generators
${\widetilde{Z}}_{\mu\nu}$ and $SO(N)$-tensorial ones ${\widetilde{Y}}_{ab}$.}

This is the expansion $osp(N|4)(2,1,2)$, which is obtained
by taking eqs.~\eqref{eazlu4.9}-\eqref{eazlu4.11} up to order two.
From \bref{eazlu4.13} and \bref{eazlu4.16} it follows that
\begin{equation}
\label{eazlu4.19}
\{ Q^a_\alpha , Q^b_\beta \}  =
  \delta^{ab} (C\gamma^\mu)_{\alpha\beta} P_\mu
  -\frac{1}{2} \delta^{ab} (C\gamma^{\mu\nu})_{\alpha\beta}
 \, {\widetilde{Z}}_{\mu\nu}  + C_{\alpha\beta}
 {\widetilde{Y}}{}^{ab} \; ,
 \end{equation}
plus
\begin{eqnarray}
\label{eazlu4.20}
[ M_{\mu\nu} , {\widetilde{Z}}_{\rho\sigma}  ]
&   =   &
 (\eta_{\rho\nu}{\widetilde{Z}}_{\mu\sigma} -
 \eta_{\sigma\nu} {\widetilde{Z}}_{\mu\rho} ) -(\mu \leftrightarrow \nu) \; ,
\nonumber\\[4pt]
[ T^{ab} , {\widetilde{Y}}^{cd}    ]
 &   = &
 (\delta^{c  b } {\widetilde{Y}}^{a  d}
- \delta^{d  b } \, {\widetilde{Y}}^{a  c}) - (a\leftrightarrow b) \; .
\end{eqnarray}
The new abelian generators (${\widetilde{Z}}_{\mu\nu} $, ${\widetilde{Y}}_{a b}$)
are the tensorial and isotensorial central charges that are added
to $N$-extended super Poincar\'{e} algebra $s{\mtP}^{(N)}$. The
$osp(N|4)(2,1,2)$ superalgebra \eqref{eazlu4.19}-\eqref{eazlu4.20}
constitutes another example of case a) at the end of Sec.~\ref{ExpGral}.
\vskip .2cm

\noindent
2) {\sl Minimal enlargement of $D=4$ $N$-superPoincar\'e algebra including $Z_{\mu \nu}$}.

Looking at \eqref{eazlu3.111}, one might simply think of making the replacement
\begin{equation}
\label{eazlu4.21}
  [ P_\mu , P_\nu ] = 0 \quad \longrightarrow \quad
 [ P_\mu , P_\nu  ] =  Z_{\mu\nu}
\end{equation}
in the $s{\mtP}^{(N)}$ algebra. Nevertheless, this would not lead to
a superalgebra since, when checking that $dd\equiv 0$ on $\omega^{\mu\nu, 4}$,
we first obtain  $d\,\omega^{\mu\nu, 4}  = -  e^{\mu , 2} \wedge e^{\nu ,2}$
({\it i.e.} the second commutator in eq.~\eqref{eazlu4.21}) and then
$dd \omega^{\mu\nu, 4}\simeq
e^{[\mu , 2} \wedge {\overline{\psi}}{}^{\alpha}_{\, a}{}^{,1} (\gamma^{\nu]})_{\alpha\beta}
\psi^{\beta}_{\, {a}}{}^{,1}\neq 0$, reflecting that the JI is not satisfied
for ($Z_{\mu \nu}, Q^a_\alpha, Q^b_\beta$). However, $dd \omega^{\mu\nu, 4}$ will
vanish if the MC equation for $d\,\omega^{\mu\nu, 4}$  is replaced (see \eqref{eazlu4.14})
by $d\,\omega^{\mu\nu, 4}  = -  e^{\mu , 2} \wedge e^{\nu ,2} +
{\overline{\psi}}{}^{\alpha}_{\, a}{}^{,1} (\gamma^{\mu\nu})_{\alpha\beta}
\psi^{\beta}_{\, a}{}^{,3}$,  which shows that an additional fermionic generator is
required. Thus, the inconsistency can be removed if the one-forms
($\omega^{\mu \nu,0}, e^{\mu,2}, \psi^{\alpha,1}_a , \omega^{\mu \nu,2}$)
are supplemented by a new fermionic one, $\psi^{\beta,3}_a$, dual to the additional
set of fermionic generators $\Sigma^i_\beta$. Then, the odd-odd sector of
the $N$-superPoincar\'e algebra is completed with a non-trivial
additional relation
\begin{equation}
\label{QE}
\{Q^a_\alpha,Q^b_\beta\}=\delta^{ab}(C\gamma^\mu)_{\alpha\beta}P_\mu \quad,\quad
\{Q^a_\alpha,\Sigma^b_\beta\}=
- \frac{1}{2}\delta^{ab}(C\gamma^{\mu\nu})_{\alpha\beta}Z_{\mu\nu} \; .
\end{equation}
This was referred to as the minimal supersymmetrization of the
Maxwell algebra in \cite{hiazlu9, hiazlu10,hiazlu11, hiazlu12}.

The new spinorial generators  $\Sigma^\beta_a$ were originally added by
Green \cite{hiazlu13} on supersting theory grounds (see further
\cite{CAIPB:00} and \cite{hiazlu4} in an expansions context) in the commutator
\begin{equation}
\label{eazlu4.24}
[  P_\mu , Q_\alpha^a ] = \gamma^\beta_{\mu \alpha} \,\Sigma_\beta^a \quad ,
\end{equation}
which here is a consequence of eq.~\bref{eazlu4.18}. Since
(eqs.~(\ref{eazlu4.9})-(\ref{eazlu4.11}))
our MC forms expansions do not contain sixth powers of $\lambda$,
we obtain as in \cite{hiazlu13} that
\begin{equation}
\label{eazlu4.26}
\{ \Sigma^a_\alpha , \Sigma^b_\beta \} = 0 \; .
\end{equation}
Eq.~\bref{eazlu4.18} for $d\psi^{\alpha,3}_a$ also gives the
commutators expressing the covariance properties of $\Sigma^a_\alpha$,
\begin{equation}
\label{eazlu4.27}
[ M^{\mu\nu} , \Sigma^a_\alpha ]  =
\frac{1}{4} (\gamma^{\mu\nu})_\alpha{}^{\beta} \, \Sigma^a_\alpha
\quad , \quad
[ T^{ab} , \Sigma^c_\alpha ]    =
 2 ( \Sigma^{a}_{\alpha}\delta^{bc} - \Sigma^{b}_{\alpha}\delta^{ac} ) \quad .
\end{equation}
\vskip .2cm

\noindent
3) {\sl $D=4$ $N$-extended Maxwell superalgebras with additional bosonic charges}

Let us write explicitly the main commutators of $osp(N|4)(4,3,4)$ that
follow from the MC equations given before.
In the expansions \eqref{eazlu4.9},  \eqref{eazlu4.10}, the
forms ($\omega^{\mu\nu,4}, e^{\mu,4}, \omega^{ab,4}$) correspond to the
highest powers in $\lambda$ and, hence, their dual generators
($Z_{\mu\nu}, Z_\mu, Y_{ab}$) are abelian. The last two modify
the $\{Q,\Sigma\}$ anticommutator of the minimal
superMaxwell algebra \bref{QE}, which becomes
\begin{equation}
\label{eazlu4.25}
\{ Q^a_{\alpha}, \Sigma^b_\beta \} = \delta^{ab}
\left[ (C\gamma^{\mu})_{\alpha\beta} \,
Z_{\mu} - \frac{1}{2}(C\gamma^{\mu\nu})_{\alpha\beta} Z_{\mu\nu}
\right] + C_{\alpha \beta} \,Y^{a b}   \; .
\end{equation}
Besides the commutators expressing the
Lorentz ($SO(N)$) covariance properties of
$\widetilde{Z}_{\mu \nu}, Z_{\mu \nu}, Z_\mu$,
($\widetilde{Y}_{a b}, Y_{ab}$), we have the non-trivial
relations
\begin{equation}
\label{eazlu4.22}
 [ {\widetilde{Y}}^{a b} , {\widetilde{Y}}^{c d} ]
=  \left(\delta^{c b } Y^{a d}
 -\delta^{d  b }Y^{a  c}\right) -(a\leftrightarrow b) \; ,
\end{equation}
\begin{equation}
\label{eazlu4.28}
[ {\widetilde{Z}}^{\mu\nu} , Q^a_\alpha ]  =   (\gamma^{\mu\nu})_\alpha{}^{\beta} \, \Sigma^a_\beta
\quad ,   \quad
[\widetilde{Y}^{ab} , Q^c_\alpha ]    =
 2 ( \Sigma^{a}_{\alpha}\delta^{bc} - \Sigma^{b}_{\alpha}\delta^{ac} ) \; .
\end{equation}

For $N=1$, the superalgebra relations  \bref{QE}-\bref{eazlu4.26}
with $Z_{\mu}=0={\widetilde{Z}}_{\mu\nu}$ were proposed in \cite{hiazlu9} as the
simplest supersymmetrization of the $D=4$ Maxwell algebra (for $N{=}1$
the generators $T_{ab}, Y_{ab}, \widetilde{Y}_{ab}$ are clearly absent).
Obviously, setting some generators equal to zero in an algebra, as done
above for $Z_\mu$ and $\widetilde{Z}_{\mu\nu}$, does not lead in general to a
subalgebra, but here all the resulting commutators satisfy the JI by virtue of the $D{=}4$
Fierz identity $(C\gamma^\mu)_{(\alpha \beta} (C\gamma_\mu)_{\gamma\delta)} = 0$,
where the bracket means symmetrization.

Other $D=4$ $N$-extended Maxwell superalgebras,  with supersymmetrized tensorial charges
${{Z}}_{\mu\nu}$, were considered  in \cite{hiazlu11,hiazlu12}.
 In \cite{hiazlu11} the following coset decomposition of $D=4$ SUSY $adS$ superalgebra
$osp(N|4)$, different from the one given by \bref{eazlu4.1},
was introduced for even $N{=}2n$,
 \begin{equation}
 \label{eazlu4.29}
 osp(2n;4)=(sl(2;C)\oplus u(n))
 \oplus
 \frac{sp(4)}{sl(2;C)} \oplus \frac{o(2n)}{u(n)} \oplus
 \frac{osp(2n;4)}{sp(4)\oplus  o(2n)}
 \end{equation}
To recover the algebras of \cite{hiazlu11} as expansions, the coset part $\frac{o(2n)}{u(n)}$
of the internal symmetry generators should be expanded in powers of
$\lambda$ in the same way as the vierbein $e^\mu$.
In \cite{hiazlu12} the $D=4$ $N$-extended Maxwell algebras were obtained as a particular
contraction of the direct sum of two real superalgebras, describing respectively
the supersymmetrization of $o(3,1)\simeq sl(2,\mathbb{C})$
($sl(k|2;\mathbb{C})$, $0\leq k \leq 2N$)
and  the supersymmetrized $o(3,2)\simeq sp(4)$ algebra
(the $D{=}4$ extended AdS  superalgebra $osp(2N-k|4)$).

\sect{Final Remarks}

The main aim of this paper was to
provide further examples showing that quite complicated (super)algebras
can be derived easily as expansions of a basic (super)algebra
which encodes some essential features (as reflected
by a certain coset decomposition). Previous studies of
the expansion method \cite{hiazlu4, hiazlu1} were
used to derive \cite{hiazlu1,hiazlu2} the $D{=}11$ full $M$-algebra
(including the Lorentz part) as a particular expansion
of $osp(1|32)$ and to look at Chern-Simons supergravities.
Also, the ($p,q$)-Poincar\'e superalgebras \cite{Ach-Tow:89} governing the
extended $D{=}3$ supergravities were found \cite{hiazlu14} to be
expansions of $osp(p+q|2)$.

Here we have applied the expansion method to
(super)algebras to obtain various Maxwell algebras
and new generalizations of the Green algebra \cite{hiazlu13}.
These (super)Maxwell algebras, all characterized by the presence of
the four-momenta commutator $[P_\mu,P_\nu]=Z_{\mu \nu}$, may be considered
as symmetries of an enlarged spacetime with additional bosonic
coordinates. Recently, it has been shown \cite{Fed-Luk:12}
that the quantization of a free particle in such a ten-dimensional
enlarged $D{=}4$ spacetime describes a Lorentz covariant extension
of the planar Landau problem (the non-relativistic particle in a
constant magnetic field background).

We have found, in particular, that the $\lambda^4$ term in
the expansion of the Lorentz MC one-forms (see eq.~\bref{eazlu3.12})
generates the tensorial `central' charges $Z_{\mu\nu}$
of the $D=4$ Maxwell algebra \bref{eazlu3.111}, which turns out
to be the expansion o(3,2)(4,2) of the $adS$ algebra $o(3,2)$.
These generators $Z_{\mu \nu}$ also appear,
as they should, in all the generalized Maxwell algebras described in
Sec.~\ref{LorMax}. The inclusion of the generators $Z_{\mu \nu}$
in a superPoincar\'e algebra to obtain a Maxwell superalgebra
requires further the addition of the Green fermionic
generator. The $N$-extended Maxwell superalgebras
are discussed in Sec.~\ref{sMax}. The most general one
considered in detail in this paper is the expansion $osp(N|4)(4,3,4)$; it
includes two sets of bosonic generators, a Lorentz vector $Z_\mu$ and the usual
$Z_{\mu \nu}$ tensor, and two other sets $\widetilde{Y}_{ab}, Y_{ab}$
that are $SO(N)$ tensors. The `minimal' $N$-extended Maxwell
superalgebra, with generators
$\{M_{\mu \nu}, P_\mu, Z_{\mu \nu}, Q^a_\alpha, \Sigma^a_\alpha, T_{ab}\}$,
may be obtained by a suitable reduction of the
general case, as shown in Sec.~\ref{sMax}.

We have not considered other (super)algebras as those of the type
given in \cite{hiazlu12} from the expansion method point of view. It is
unclear whether they can be obtained by this procedure,
but we recall here that contractions do appear as a particular case of
expansions \cite{hiazlu1} (see also \cite{hiazlu4,hiazlu2,hiazlu14}).

\vskip .5cm
\noindent
{\bf Acknowledgments.} This paper has been supported by research grants from the
Spanish MINECO (FIS2008-01980, FIS2009-09002, CONSOLIDER
CPAN-CSD2007-00042), from the Polish Ministry of Science and
Education (202332139) and from the Polish National Science Center
(project 2011/01/B/ST2/0335).


\begin{thebibliography}{20}

\bibitem{hiazlu3}
J.A.~de Azc\'arraga, J.M.~Izquierdo,
{\it Lie groups, Lie algebras, cohomology and some applications in physics},
Camb. Univ. Press, 1995.

\bibitem{hiazlu4}
M.~Hatsuda and M.~Sakaguchi,
Progr. Theor. Phys. {\bf 109}, 853-867 (2003)
[hep-th/0106114].

\bibitem{hiazlu1}
J.A.~de Azc\'arraga, J.M.~Izquierdo, M.~Pic\'{o}n, O.~Varela,
Nucl. Phys. {\bf B662}, 185-219 (2003)
[hep-th/0212347].

\bibitem{hiazlu2}
J.A.~de Azc\'arraga, J.M.~Izquierdo, M.~Pic\'{o}n, O.~Varela,
Class. Quant. Grav. {\bf 21}, S1375-S1384 (2004)
[hep-th/0401033];
Int.\ J.\ Theor.\ Phys.\  {\bf 46} (2007) 2738-2752 (2007)
[hep-th/0703017 [hep-th]].

\bibitem{IRP:06}
F.~Izaurieta, E.~Rodr\'iguez and P.~Salgado,
J. Math. Phys.  {\bf 47}, 123512-1-28 (2006)
[hep-th/0606215].

\bibitem{W-W:00}
E. Weimar-Woods, J. Math. Phys. {\bf 36}, 4519-4548 (1995);
Rev. Math. Phys. {\bf 12}, 1505-1529 (2000).

\bibitem{hiazlu5}
H.~Bacry, P.~Combe, J.L.~Richard,
Nuovo Cim. {\bf A67},  267-299 (1970).

\bibitem{hiazlu6}
R.~Schrader, Fortschr. Phys. {\bf 20}, 701-734 (1972).

\bibitem{Ne-Ol:00}
J. Negro and M.A. del Olmo,
J. Math. Phys {\bf 31}, 568-578, 2811-2821 (1990).

\bibitem{hiazlu7}
S.~Bonanos, J.~Gomis, J. Phys. {\bf A43}, 015201 (2010)
 [arXiv:0812.4140 [hep-th]].

\bibitem{hiazlu8}
J.~Gomis, K.~Kamimura, J.~Lukierski, JHEP08, 039 (2009).
[arXiv:1005.3714 [hep-th]].

\bibitem{hiazlu9}
S.~Bonanos, J.~Gomis, K.~Kamimura, J.~Lukierski,
Phys. Rev. Lett. {\bf 104}, 090401 (2010)
[arXiv:0911.5072 [hep-th]].

\bibitem{hiazlu10}
S.~Bonanos, J.~Gomis, K.~Kamimura, J.~Lukierski,
J. Math. Phys. {\bf 51}, 102301 (2010)
 [arXiv:1005.3714 [hep-th]].

\bibitem{hiazlu11}
J.~Lukierski,
Proc. Stekl. Inst. Math. {\bf 272}, 1-8 (2011)
[arXiv:1007.3405 [hep-th]].

\bibitem{hiazlu12}
 K.~Kamimura, J.~Lukierski,
 Phys.\ Lett.\ B {\bf 707}, 292-297 (2012)
 [arXiv:1111.3598 [math-ph]].

\bibitem{hiazlu13}
M.~B.~Green,
Phys.\ Lett.\ B {\bf 223}, 157-164 (1989).

\bibitem{CAIPB:00}
C.~Chryssomalakos, J.~A.~de Azc\'arraga, J.~M.~Izquierdo and J.~C.~P\'erez Bueno,
Nucl. Phys. {\bf B567}, 293 (2000)
[hep-th/9904137].

\bibitem{Ach-Tow:89}
A. Ach\'ucarro and P.K. Townsend,
Phys. Lett. {B 229}, 383-387 (1989).

\bibitem{hiazlu14}
J.~A.~de Azc\'arraga and J.~M.~Izquierdo,
Nucl. Phys. {\bf B854}, 276-291 (2012)
[arXiv:1107.2569 [hep-th]].

\bibitem{Fed-Luk:12}
S.~Fedoruk and J.~Lukierski,
Phys. Lett. {\bf B718}, 646-652 (2012)
[arXiv:1207.5683 [hep-th]].

\end{thebibliography}
\end{document}